\documentclass[twocolumn,showpacs,prl]{revtex4}
\def\rf#1{(\ref{eq:#1})}
\def\lab#1{\label{eq:#1}}

\def\br{\begin{eqnarray}}
\def\er{\end{eqnarray}}
\def\be{\begin{equation}}
\def\ee{\end{equation}}
\def\({\left(}
\def\){\right)}
\def\u2{\mid u\mid^2}
\def\rlx{\relax\leavevmode}
\def\IR{\rlx\hbox{\rm I\kern-.18em R}}

\input{epsf}

\usepackage{graphics}

\begin{document}


\title{Exact time dependent Hopf solitons in $3+1$ dimensions}

\author{L. A. Ferreira}
\affiliation{Instituto de F\'\i sica de S\~ao Carlos; IFSC/USP;\\ 
Av. Trabalhador S\~ao Carlense 400; CEP 13560-970; 
S\~ao Carlos-SP, Brazil\\
and\\
Instituto de F\'\i sica Te\'orica, IFT/UNESP; \\
Rua Pamplona 145, CEP 01405-900, S\~ao Paulo-SP, Brazil}

\date{\today}

\begin{abstract}
We construct an infinite number of exact time dependent soliton
solutions, carrying non-trivial Hopf topological charges, in a $3+1$
dimensional Lorentz invariant theory with target space $S^2$. The
construction is based on an ansatz which explores the invariance of
the model under the conformal group $SO(4,2)$ and the
infinite dimensional group of area preserving diffeomorphisms of
$S^2$. The model is a rare example of an integrable theory in four
dimensions, and the solitons may play a role in the low energy limit of gauge
theories.  
\end{abstract}


\maketitle

\noindent{\em {\underline{Introduction}} - } Solitons have a very
important role in several areas of Physics. In particular,  they are
useful in the understanding of 
many non-perturbative (strong coupling) phenomena. The appearance of
solitons requires a rich  
symmetry structure, leading to conservation laws, and therefore they
 underlie the integrability properties of the models. Although the
$1+1$ soliton theory is well developed, very few exact results are
known about solitons in higher dimensions \cite{afsg}. In this Letter
we construct an infinite number of exact soliton solutions, carrying
Hopf topological charges, in a $3+1$ dimensional, Lorentz invariant
theory, and possesing an infinite number of conservation laws. An
important feature of the solitons is that, in the far past and far
future, their energy density is vanishingly 
small and distributed over a large region of space. For finite times
the energy density builds up in a small portion of space. In addition,
the solutions depend upon a free parameter that allows to re-scale
their size and rate of time evolution. The
condition for the total energy to be conserved also implies that the Hopf
topological charge should be non-trivial. The model is a rare and interesting
example of an integrable theory in four dimensions, and its solitons may
have an important role in the low energy limit of the Yang-Mills
theories, as we discuss at the end of this Letter. In addition, our
work may be of interest in the study of the solitons of the
Skyrme-Faddeev model \cite{fn}. 
The theory is defined by the action 
\be
S=  - \frac{1}{e^2} \int  d^4x \, H_{\mu\nu}^2
\lab{action}
\ee
where 
$H_{\mu\nu}$ is the pull-back of the area form on $S^2$
\be
H_{\mu\nu}\equiv 
-2i\frac{\(\partial_{\mu} u\partial_{\nu} u^* - 
 \partial_{\nu} u \partial_{\mu} u^*\)}{\(1+\u2\)^2}  
=  {\vec n}\cdot\(\partial_{\mu}{\vec n} \wedge 
\partial_{\nu}{\vec n}\)
\lab{hdef} 
\ee
with $u$ being a  complex scalar field, related to the triplet of
scalar fields
${\vec n}$ living on $S^2$ 
(${\vec n}^2=1$) through the stereographic projection 
${\vec n} = \(u+u^*,-i\(u-u^*\),\u2 -1\)/\(1+\u2\)$. 
The Euler-Lagrange equations are
\be
\partial_{\mu}{\cal K}^{\mu}=0 \; ;\qquad\qquad 
{\cal K}_{\mu}\equiv H_{\mu\nu}\, \partial^{\nu} u
\lab{eqmotion}
\ee
together with its complex conjugate. The action \rf{action} and the
eqs. of motion \rf{eqmotion} are invariant under the conformal group
$SO(4,2)$ of four-dimensional Minkowski space-time
\cite{babelon}. They are also 
invariant under the 
area preserving diffeomorphisms of $S^2$, and  the infinite set of
associated Noether currents are given by \cite{razumov}
\be
J_{\mu}^G = \(\delta G/\delta u\) {\cal K}_{\mu} +   
\(\delta G/\delta u^*\) {\cal K}_{\mu}^*
\lab{noethercurr}
\ee
where $G$ is any functional of $u$ and $u^*$, but not of their
derivatives. 

\noindent{\em {\underline{Solutions}} -} We introduce the
  coordinates\footnote{Notice they correspond to the 
  toroidal coordinates of ref \cite{afz99} for $\zeta=0$ and
  $y=1/\sinh^2\eta$.} 
\br
x^0&=& \(a/p\)\, \sin \zeta \; ; \quad \;\;\,
x^1= \(a/p\)\, \cos \varphi/\sqrt{1+y}\lab{minktoro}\\  
x^3&=& \(a/p\) \sin \xi  \sqrt{y/\(1+y\)};\; \,
x^2= \(a/p\) \sin \varphi/\sqrt{1+y}
\nonumber
\er
with $p=\cos \zeta - \cos \xi \, \sqrt{y/(1+y)}$, and $a$ is a constant
with dimension of length. The range of the coordinates are: $y\geq 0$,
$0\leq \xi\, , \, \varphi \leq 2\pi$, and $0\leq \zeta \leq
\pi$. Notice that the range of $\zeta$ is restricted because $\( \zeta
, y , \xi , \varphi\)$ and $\( \zeta+\pi, y , \xi+\pi , \varphi+\pi\)$
give the same point on Minkowski space-time. We introduce the ansatz
  \cite{afz99,babelon} 
\be
u = \sqrt{\(1-g\)/g}\;\; e^{i\(m_1\,\xi+m_2\, \varphi+m_3\,\zeta\)}
\lab{ansatz}
\ee
with $g = g\(y\)$, and $0\leq g \leq 1$. In order for $u$ to be single
valued we need $m_1$ and $m_2$ to be integers. In addition, $\( \zeta=0
, y , \xi , \varphi\)$ and $\( \zeta=\pi, y , \xi+\pi , \varphi+\pi\)$
correspond to the same point $\(x^0=0, x^1,x^2,x^3\)$. Therefore, we
also need $m_1+m_2+m_3= 2 N$, 
with $N$ being an integer, in order for $u$ to be
single valued.
Replacing \rf{ansatz} into \rf{eqmotion} we reduce those  
four-dimensional non-linear partial differential equations into a
single linear ordinary differential equation given by
\be
\partial_y\( \Lambda \, \partial_y g\) = 0\; ; \quad 
\Lambda \equiv m_1^2\,\(1+y\)+m_2^2\, y\(1+y\)-m_3^2\, y
\lab{eqforg}
\ee
The analysis of the solutions is very simple. For $m_2=0$ the solution
  is logarithmic and so diverges for $y\rightarrow \infty$. Since we
  need $0\leq g \leq 1$, for $y\geq 0$, the only acceptable solution
  is $g={\rm constant}$, which we shall discard. For the same
  reasons $\Lambda$ can not have real and positive zeroes for 
  $m_2\neq 0$. Since those are given by  $y_{\pm}=-b\pm\sqrt{\Delta}$, with 
\br
b&=&\left[\(m_1+m_3\)\(m_1-m_3\)+m_2^2\right]/2m_2^2 
\lab{bdeltadef}\\
\Delta&=&
\left[\(m_1+m_3\)^2-m_2^2\right]\left[\(m_1-m_3\)^2-m_2^2\right]/4m_2^4
\nonumber 
\er
we can not have $b<0$ and $\Delta \geq 0$, which happen whenever
  $\(m_1+m_3\)/m_2 \geq 1$ and $\(m_1-m_3\)/m_2 
\leq -1$ or $\(m_1-m_3\)/m_2 \geq 1$ and $\(m_1+m_3\)/m_2
\leq -1$. Therefore, the  
solutions satisfying the boundary conditions $g\(0\)=1$ and
$g\(\infty\)=0$, are ($m_2\neq 0$)\footnote{The ArcTan is assumed to
  take values from $0$ to $\pi$.} 
\br
g&=& \frac{b}{y+b} \; ; \qquad {\rm for} \;\; \Delta=0 \; ; \; \; \; 
b>0
\lab{normsolg}\\
g&=& \frac{{\rm ArcTan} \(\sqrt{-\Delta}/\(y+b\)\)}{{\rm ArcTan}
  \(\sqrt{-\Delta}/b\)} \; ; \quad {\rm for} \;\; \Delta<0 \; ; \; \; \; 
{\rm any} \;\; b\nonumber\\
g&=&
\frac{\ln\left[\(y+b+\sqrt{\Delta}\)/\(y+b-\sqrt{\Delta}\)\right]}{
\ln\left[\(b+\sqrt{\Delta}\)/\(b-\sqrt{\Delta}\)\right]}
\; ; \;\; {\rm for} \;\; \Delta,b>0 \nonumber
\er
They are all monotonically decreasing functions of $y$,
from $1$ at $y=0$, to $0$ for $y\rightarrow \infty$. 
  
\begin{figure}
\scalebox{.75}{
\includegraphics{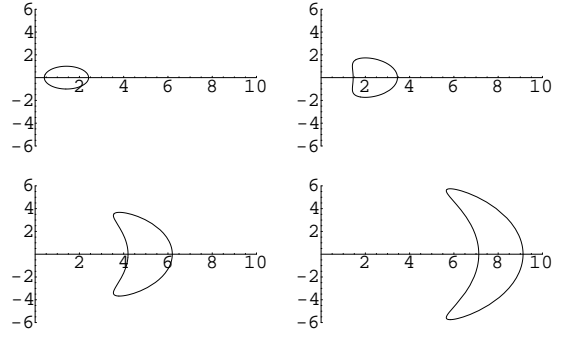}}
\caption{\label{fig:crosssec} Cross sections of the surfaces of
	constant $n_3$ for 
	$n_3=1-2g\(y=1\)$, and at the times $\tau =ct/a=0,2,5,8$. The
	vertical and horizontal axis correspond to $x^3/a$ and
	$\rho/a$ respectively. The surfaces are
	invariant under $\tau\rightarrow -\tau$     }
\end{figure}

In order to visualize the time evolution of the solutions we have to
take slices of constant $x^0=c\, t$. The best way to do it, is to trade the
coordinate $\zeta$ in favour of the dimensionless time $\tau =
ct/a$. From \rf{minktoro} one has that $\tau^2 p^2= 1-\cos^2 \zeta$,
which is a quadratic equation for $\cos \zeta$ in terms of
$\tau^2$. The two solutions lead to equivalent descriptions of the
constant time slices. With one of those choices, the Cartesian space
coordinates on the time slices are written as 
\be
x^1 = \frac{a}{q}\, \cos \varphi\; ; \quad 
x^2= \frac{a}{q}\, \sin \varphi\; ; \quad 
x^3 = \frac{a}{q}\, \sin \xi \, \sqrt{y}
\lab{constanttimeslice}
\ee
with $q=\frac{\sqrt{1+y+\tau^2\(1+y\sin^2\xi\)}-\cos\xi\sqrt{y}}{
1+\tau^2}$. 
The form of the solutions can be understood through their 
surfaces of constant $n_3$, the third component of the scalar fields
on $S^2$ (see \rf{hdef} and below). From \rf{ansatz} one has
that $n_3=1-2g$ . So, fixed $n_3$ means fixed
$g$, which in its turn means fixed $y$, since from
\rf{normsolg}, we have that $g$ is a monotonic function of
$y$. Therefore, for a given fixed time $\tau$ the surfaces of constant
$n_3$ are obtained from \rf{constanttimeslice}  by varying the angles
$\xi$ and $\varphi$, and keeping $y$ fixed. Notice that such surfaces
are valid for any solution given in \rf{ansatz} and \rf{normsolg}. The
only thing is that the chosen fixed value of $y$ corresponds to
different values of $n_3$ for different solutions. From
\rf{constanttimeslice} and the form of $q$, we see that such surfaces
are invariant under rotations around the $x_3$-axis, and under the
time reflection $\tau \rightarrow -\tau$. They are toroidal surfaces
around the $x^3$-axis. In
Figure \ref{fig:crosssec} we show their cross sections, at some fixes
values of time, through the half-plane $x^3\rho$, with
$\rho=\sqrt{x_1^2+x_2^2}$.  Some general
properties of the solutions are: {\em i)} The surface for $n_3=-1$,
which implies $g=1$ and so $y=0$, is a circle on the plane $x_3=0$
with center at the origin and radius $a\sqrt{1+\tau^2}$; {\em ii)} The
surface for $n_3=1$, which implies $g=0$ and so $y\rightarrow \infty$,
corresponds to the $x_3$-axis plus the spatial infinity, for any time
$\tau$ (The reason is
that for $\xi\neq 0$, the limit $y\rightarrow \infty$ gives $\mid
q\mid \rightarrow \infty$, and for $\xi =0$ gives $q\rightarrow 0$);
{\em iii)} For $\tau =0$ the surfaces of constant $n_3$, for $-1 < n_3
< 1$, are torus centered around the origin with a tickness that grows
as $n_3$ varies from $-1$ to $1$. As $\tau$ flows towards the future
or the past, those torus get ticker and their cross section deform 
from a circle to a quarter moon shape as shown in Figure
\ref{fig:crosssec}; {\em iv)} The 
solution performs one single oscillation as $\tau$ varies from
$-\infty$ to $\infty$.  Although the surfaces of constant $n_3$ are
symmetrical under  the interchange $\tau
\leftrightarrow -\tau$, the same is not true for the energy density as
shown below; {\em v)} Due to Derrick's scaling
  argument, the theory \rf{action} can not have stable static
  solutions in $3+1$ dimensions, and so our solutions can not be put
  at rest, even though the
  limit of large $a$ slow down their time evolution.

\noindent{\em {\underline{Hopf charge}} - } The condition for finite
energy requires the field ${\vec n}$ to be 
constant at spatial infinity. For any fixed time, our solutions have 
${\vec n}\rightarrow \(0,0,1\)$ for $r\rightarrow \infty$
($r^2=x_1^2+x_2^2+x_3^2$). Therefore, for topological considerations, 
we can compactify $\IR^3$ into $S^3$, and the solution defines a Hopf
map $S^3\rightarrow S^2$, for any fixed time. The Hopf index is
calculated as follows: given the solution \rf{ansatz} and
\rf{normsolg} we map $\IR^3$ into $S^3_Z$ through
$
Z=
\( 
\begin{array}{c}
z_1\\
z_2
\end{array}\)=
\( 
\begin{array}{c}
\sqrt{1-g}\, e^{i\(m_1 \xi+m_3\zeta\)}\\
\sqrt{g}\, e^{-i\(m_2\varphi\)}
\end{array}\)
$
with $Z^{\dagger}\, Z=1$, 
and so the four real parameters in $Z$ parametrize $S^3_Z$. Then we
map $S^3_Z$ into $S^2_u$ through $u=z_1/z_2$. The Hopf index is
defined as 
\be
Q_H= \frac{1}{4\pi^2}\, \int d^3x\, {\vec A}\cdot\(\nabla \wedge {\vec
  A}\)
\lab{hopfdef}
\ee
with
${\vec A}=i\(Z^{\dagger}{\vec \nabla} Z- {\vec \nabla} Z^{\dagger}\,
Z\)/2$. Using \footnote{From  
  \rf{minktoro}: $y=\frac{\(a^2+s^2\)^2+4a^2x_3^2}{4a^2\rho^2}$,  
$\tan \varphi = \frac{x^2}{x^1}$, $\tan \zeta=\frac{2ax^0}{a^2-s^2}$, and 
$\tan\xi=-\frac{2ax^3}{a^2+s^2}$,  with
  $s^2=x_0^2-\rho^2-x_3^2$.}, one  then gets   
${\vec A}\cdot\({\vec \nabla}\wedge {\vec A}\)= 
-\frac{\partial_y g}{a\rho^4} m_2 \( m_1 
\( a^2+R^2\) +2 m_3 \, x^0\, x^3\)$, 
with $R^2=x_0^2+x_1^2+x_2^2+x_3^2$.   
Notice that  $y$ is an even function of
all $x^{\mu}$'s [12] and so is $\partial_y
g$. Therefore, the term proportional to $m_3$, being odd in $x^3$,
vanishes when integrated on space. 
The volume element on the time slices  \rf{constanttimeslice} is 
\be
d^3 x  =  dy\, d\xi \, d\varphi\, a \rho^4/\(a^2+R^2\)
\lab{volumeconsttime}
\ee
The Hopf index \rf{hopfdef} for the 
solutions \rf{ansatz} and \rf{normsolg} is then   
\be
Q_H= -m_1m_2\(g\(\infty\)-g\(0\)\) = m_1m_2
\lab{hopfchargefinal}
\ee  
\noindent{\em {\underline{Noether charges}} - } Among the charges
associated to the currents \rf{noethercurr} there is an infinite
abelian subset corresponding to the cases where $G$ is
a functional of the norm of $u$ only, or equivalently a functional of
$g=1/(1+\mid u\mid^2)$ (see \rf{ansatz}). One can easily check that
the Poisson brackets of the densities $J_0^G$, associated to such
choice of $G$, does vanish \cite{razumov}. If one substitute
\rf{ansatz} into \rf{noethercurr} and uses [12] one gets
that, for $G$ being a functional of $g$ only, 
$J_0^G=4\(\partial_y g\)^2 \frac{\delta G}{\delta g}\left[2 m_1 \(1+y\) 
x^0x^3+m_3 y \(a^2+R^2\)\right]/a\rho^4$. 
Since $y$ is an even function of all Cartesian coordinates
$x^{\mu}$, it follows that  the term propotional to $m_1$, being
odd in $x^3$,  vanishes when integrated on the whole space. Using
\rf{volumeconsttime} one gets that the corresponding Noether charges
are $Q^G= 16\pi^2m_3\int_0^{\infty}dy\, y\, \(\partial_y g\)^2 
\frac{\delta G}{\delta g}$. 
Using \rf{eqforg} and \rf{normsolg} one gets 
\be
\partial_y g = -\,\frac{C}{\Lambda} \; ; \quad 
C =\(m_1^2+m_2^2-m_3^2\)w/\ln\frac{1+w}{1-w}
\lab{derg}
\ee
with $w=\sqrt{\Delta}/b$, and $\Delta$, $b$ satisfying any of the
three conditions in 
\rf{normsolg}. Notice that in all those cases we have $C>0$, and in
particular for $\Delta =0$ one has $C=\(m_1^2+m_2^2-m_3^2\)/2$.   
Then taking $G=g^n/16\pi^2n!$, one gets that such
charges evaluated on the solutions \rf{normsolg} are given by
\be
Q^{(n)}= m_3\; F^{(n)}\(w\)\; ; \qquad  n=1,2,3\ldots
\lab{qnoethern}
\ee
with
\br
 F^{(n)}\(w\)=
\(\ln (1+w)/(1-w)\)^{-n-1}\left[ -2
  \epsilon_{-}\(n\) \right.
\nonumber\\
+    
\sum_{l=1}^n \left. \(\frac{\epsilon_{+}\(n-l\)}{w}-\epsilon_{-}\(n-l\)\) 
\frac{1}{l!}
\(\ln\frac{1+w}{1-w}\)^l\right]
\lab{fndef}
\er
with $\epsilon_{\pm}\(n\)=\(1\pm\(-1\)^n\)/2$. For the case $\Delta=0$,
those charges simplify to $Q^{(n)}= m_3 n/(n+2)!$. The case $n=1$
corresponds to the $U(1)$ subgroup of the $S^2$-area preserving
diffeomorphism group generated by $u \rightarrow e^{i\alpha} u$.  

\begin{figure*}
\scalebox{1.395}{
\includegraphics{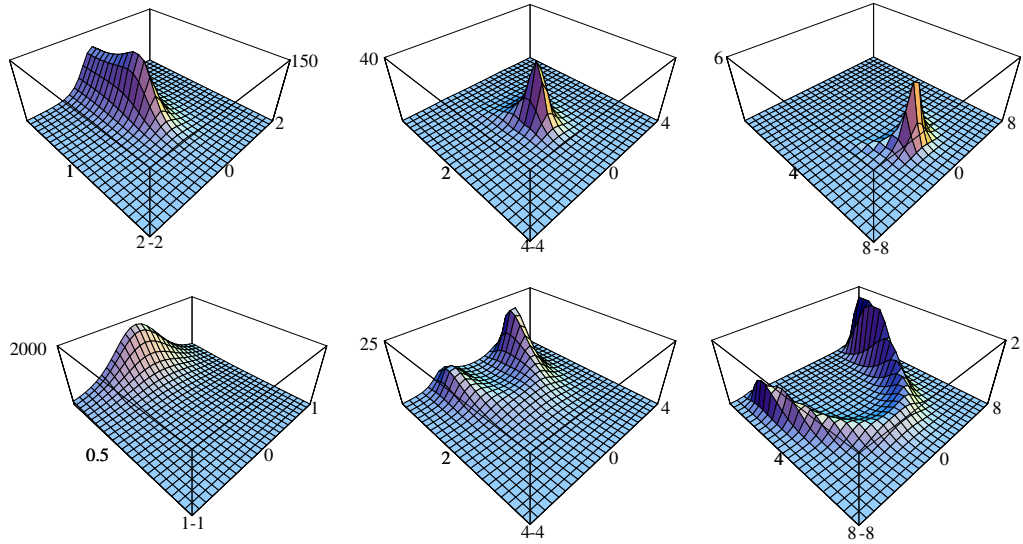}}
\caption{\label{fig:full} Plots of the energy density ${\cal H}$
  \rf{energydensity}, in units of $\frac{8}{e^2}$, as a function of
  $\rho/a$ (left-front axis) and $x^3/a$ (right-front axis). ${\cal
  H}$ is invariant under 
  rotations around the $x^3$-axis and under the joint parity 
  transformations $\tau\rightarrow -\tau$ and $x^3\rightarrow
  -x^3$. The top row correspond to the 
  soliton with $\(m_1,m_2,m_3\)=\(1,4,1\)$, at the times $\tau =ct/a=
  0,2,6$, and the bottom row to the soliton with
  $\(m_1,m_2,m_3\)=\(4,1,1\)$, at the same times. These two solitons
  have the same total energy \rf{energyfinal}, Noether charges
  \rf{qnoethern} and Hopf charge \rf{hopfchargefinal}, but the soliton
  $\(1,4,1\)$ has four times more angular momentum \rf{angmom} than the soliton
  $\(4,1,1\)$. } 
\end{figure*}

\noindent{\em {\underline{Angular Momentum}} - } The angular momentum
is given by $L_i=\frac{1}{2}\varepsilon_{ijk}\int d^3x M^0_{(jk)}$,
with $M^{\mu}_{(\rho\sigma)}=T^{\mu}_{\rho}x_{\sigma} -
T^{\mu}_{\sigma}x_{\rho}$, and $T^{\mu}_{\nu}$ being the canonical
energy-momentum tensor associated to \rf{action}. For the
solutions \rf{ansatz} and \rf{normsolg} it is   
\be
L_3=\frac{128\pi^2}{e^2}\, m_2\, m_3 \, F^{(1)}\(w\) \, ; \qquad
\;\;L_1=L_2=0
\lab{angmom}
\ee
\noindent{\em {\underline{Energy}} - } The Hamiltonian density
associated to \rf{action} is ${\cal H}=(2/e^2)\(\sum_i
H_{0i}^2+\sum_{i<j} H_{ij}^2\)$, with $i,j=1,2,3$. For the ansatz
configurations \rf{ansatz} one gets that
\be
{\cal H}=\frac{8}{e^2}\(\partial_y g\)^2\left( m_1^2 {\cal E}_1 +m_2^2
  {\cal E}_2 +   
m_3^2 {\cal E}_3 + 2 m_1 m_3 {\cal E}_{13}\right) 
\lab{energydensity}
\ee
with 
${\cal E}_{13} = \frac{4x^0x^3\(a^2+R^2\)}{a^2\rho^6}$, 
${\cal E}_1 = \frac{4}{\rho^4}\( 1+y +
\frac{2x_0^2r^2}{a^2\rho^2}\)$, 
$ {\cal E}_2 = \frac{4}{\rho^4}\( y\(1+y\) + 
\frac{x_0^2\(a^2+s^2\)^2}{2a^4\rho^2}\)$, and 
${\cal E}_3 = \frac{4}{\rho^4}\( y +
\frac{2x_0^2x_3^2}{a^2\rho^2}\)$. 
The energy density is axially symmetric, and invariant under the
joint parity 
transformations $x^0\rightarrow -x^0$ and $x^3\rightarrow -x^3$. In
Figure \ref{fig:full} we show the time evolution of ${\cal H}$ for two
particular solitons. Notice that it resembles  the time evolution
of some types of Ward's solitons \cite{ward}.

The total energy is only conserved when $m_1,m_2\neq  0$, and so when
the Hopf charge \rf{hopfchargefinal} is non vanishing. Indeed,  the
integration of the term involving ${\cal E}_{13}$ vanishes since it is
odd in $x^3$. The other terms give 
$\int d^3x \(\partial_y g\)^2 {\cal E}_3 =
\frac{8\pi^2}{a}\int_0^{\infty} dy \,\(\partial_y g\)^2 \, y$, and 
$\int d^3x \(\partial_y g\)^2 \( m_1^2 {\cal E}_1+m_2^2 {\cal E}_2\) =
\frac{8\pi^2}{a}\left[ \frac{\tau^2}{1+\tau^2}
W 
+ \int_0^{\infty} dy \,\(\partial_y g\)^2 \,
\( m_1^2 \, \(1+y\)+ m_2^2 \, y\(1+y\)\) \right]$, 
with $W=\int_0^{\infty} dy \,\(\partial_y g\)^2 \,\(m_1^2-m_2^2 \,
y^2\)$. From \rf{derg} one then gets that
$W=C^2y/\Lambda\mid_{y=0}^{y=\infty}$. In our analysis below
\rf{bdeltadef} we concluded that we have to have $m_2\neq 0$ for the
solutions to be well behaved. But $W$ does not vanish if $m_1=0$ and
$m_2\neq 0$. Therefore, in order for the total energy to be time
independent we need $m_1\neq 0$ too. Consequently, 
$E=\frac{\(8\pi\)^2}{a \, e^2}\int_0^{\infty} dy \, \(\partial_y g\)^2
\,  \Omega$, 
with $\Omega= m_1^2 \(1+y\) + m_2^2\,  y\(1+y\) +  m_3^2\,  y = \Lambda +
2 m_3^2 \, y $. Using \rf{derg} one gets $\int_0^{\infty} dy \,
  \(\partial_y g\)^2 \Lambda = - C \( g\(\infty\)-g\(0\)\)$, and the
  remaining integral is proportional to $Q^{(1)}$ (see
  \rf{qnoethern}). Then one gets that   
\be
E=\frac{\(8\pi\)^2}{a \, e^2}\left[ C + 2\, m_3^2\, F^{(1)}\right]
\lab{energyfinal}
\ee
with $C$ and $F^{(1)}$ given in \rf{derg} and \rf{fndef}. 
In the case $\Delta=0$, which implies $m_2^2=\(m_1\pm m_3\)^2$ (see
\rf{bdeltadef}), the energy reduces to $E=\frac{\(8\pi\)^2}{a \,
  e^2}\( m_1^2 +\frac{1}{3}m_3^2\pm m_1m_3\)$. 
The energy $E$ \rf{energyfinal} is invariant under the interchange
$m_1\leftrightarrow m_2$, and under the change of sign of
any integer $m_i$, $i=1,2,3$, individually. Therefore, it is
$16$-fold degenerate for $m_1\neq m_2$, $m_3\neq 0$, and that is
reduced by factors 
$2$'s according $m_1=m_2$ or $m_3=0$ (remember
there are no physically acceptable solutions for $m_1,m_2=0$). In any
case, such degeneracy is completely lifted by considering the values of
the Hopf charge $Q_H$ \rf{hopfchargefinal}, the 
Noether charge $Q^{(1)}$ \rf{qnoethern} (or any other $Q^{(n)}$), and
the angular momentum $L_3$ \rf{angmom}. 

The solitons we have constructed have a connection with the
Yang-Mills (YM) theory. At the classical level, they correspond to vacuum
configurations of YM. In fact, any solution of any field theory with
target space $S^2$ can be mapped into a vacuum of $SU(2)$  YM. Indeed, consider
a $SU(2)$ 
gauge theory with gauge potencial ${\vec W}_{\mu}$, with a Higgs field
${\vec \phi}$ in the triplet representation (the arrows stand for the
orientation in the $SU(2)$ algebra). The Higgs vacuum, 
$V\(\phi\)=0$
and $D_{\mu}{\vec \phi}=0$, is achieved with ${\vec \phi}=v {\vec n}$, and
${\vec W}_{\mu}= \frac{1}{e}{\vec n}\wedge \partial_{\mu}{\vec n} +
{\vec n} B_{\mu}$, where $e$ is the gauge coupling constant, ${\vec
  n}^2=1$, $v$ is the minimum of $V$, 
and $B_{\mu}$ is an arbitrary $U(1)$ gauge potential. That is in fact,
the field 
configuration of a 't Hooft-Polyakov monopole away from its core. The
field strength is ${\vec F}_{\mu\nu}={\vec n}\(\frac{1}{e} H_{\mu\nu}+
\partial_{\mu}B_{\nu}-\partial_{\nu}B_{\mu}\)$, with $H_{\mu\nu}$ as
in \rf{hdef}. If we take $B_{\mu}$ to
be proportional to the potential of the Hopf charge density
\rf{hopfdef}, i.e. $B_{\mu} = -\frac{i}{e}\(Z^{\dagger}\partial_{\mu}Z-
\partial_{\mu}Z^{\dagger}\, Z\)$, then
$\partial_{\mu}B_{\nu}-\partial_{\nu}B_{\mu} = -\frac{1}{e}
H_{\mu\nu}$, for any $Z$, and  ${\vec F}_{\mu\nu}$ vanishes. Notice
that, although ${\vec F}_{\mu\nu}$ is local in the fields ${\vec n}$, 
the same is not true for ${\vec W}_{\mu}$. The same connection can be
made with a pure YM (without Higgs) using the Cho-Faddeev-Niemi
decomposition of ${\vec W}_{\mu}$ \cite{fndecomp}. Such connection with YM
is independent of the dynamics of the fields ${\vec n}$. The theory
\rf{action} becomes relevant at the quantum level. At low energies it
is reasonable to take ${\vec n}$ as an order parameter and so a degree
of freedom of ${\vec W}_{\mu}$. The low
energy effective action $S_{\rm eff.}$ of YM will then contain
\rf{action} as one of its terms, since it is a marginal operator
\cite{gies}. The  
classical solutions of $S_{\rm eff.}$ play an important role in the
generating functional calculations, and the solutions we
have constructed can then perhaps be useful in a perturbative 
expansion of $S_{\rm eff.}$ around \rf{action}. See \cite{dubois} for
a further connection between \rf{action} and YM. 

{\bf Acknowledgements:} The author is grateful to J.~Sanchez Guillen
and W.~J.~Zakrzewski for helpful discussions. Work partially
supported by CNPq.

\end{document}